\documentclass[sigconf]{acmart}
    \usepackage[utf8]{inputenc}
    \usepackage{tikz}
    \usepackage[bold]{hhtensor}
    \usepackage{balance}
    \usepackage{tabularx}
    \usepackage{verbatimbox}

\copyrightyear{2025} 
\acmYear{2025} 
\setcopyright{none}
\renewcommand\footnotetextcopyrightpermission[1]{}

\begin{document}

\fancyhead{}
\title{Teaching Quantum Computing through Lab-Integrated Learning: Bridging Conceptual and Computational Understanding}

\author{Umar Farooq}
\affiliation{%
  \institution{Louisiana State University}
  \city{Baton Rouge}
  \state{LA}
  \country{USA}}
\email{ufarooq@lsu.edu}

\author{Krishna Upadhyay}
\affiliation{%
  \institution{Louisiana State University}
  \city{Baton Rouge}
  \state{LA}
  \country{USA}}
\email{kupadh4@lsu.edu}


\begin{abstract}
Quantum computing education requires students to move beyond classical programming intuitions related to state, determinism, and debugging, and to develop reasoning skills grounded in probability, measurement, and interference. This paper reports on the design and delivery of a combined undergraduate and graduate course that employed a lab-integrated learning model to support conceptual change and progressive understanding. The course paired lectures with weekly programming labs that served as environments for experimentation and reflection. These labs enabled students to confront misconceptions and refine their mental models through direct observation and evidence-based reasoning.
Instruction began with Quantum Without Linear Algebra (QWLA), which introduced core concepts such as superposition and entanglement through intuitive, dictionary representations. The course then transitioned to IBM Qiskit, which provided a professional framework for circuit design, noise simulation, and algorithm implementation. Analysis of student work and feedback indicated that hands-on experimentation improved confidence, conceptual clarity, and fluency across representations. At the same time, it revealed persistent challenges in debugging, reasoning about measurement, and understanding probabilistic outcomes.
Our findings highlight which quantum concepts benefit most from lab-based experimentation and where programming support remains insufficient.
\end{abstract}

\maketitle

\section{Introduction}

Quantum computing (QC) is redefining the limits of computation by applying quantum-mechanical principles such as superposition, entanglement, and interference to process information in fundamentally new ways. These principles enable algorithms that can outperform classical methods on problems such as integer factorization~\cite{shor}, unstructured search~\cite{grover}, and molecular simulation~\cite{nwchem}. As both research and industry accelerate progress toward scalable quantum hardware, the demand for a workforce proficient in both the theory and practice of quantum computing is growing rapidly.

Meeting this demand presents a distinct challenge for computer science education. Most existing quantum computing courses are offered at the graduate level and focus heavily on mathematical derivations or the physical implementation of qubits. While these topics provide theoretical depth, they often limit accessibility for students with backgrounds in programming and algorithmic reasoning. On the other hand, courses that focus solely on software may encourage superficial reproduction of circuits without fostering conceptual understanding. Addressing this gap requires pedagogical approaches that integrate experimentation with conceptual learning and link abstract ideas to computational representations.

At the Louisiana State University~(LSU), we designed and taught a course titled Introduction to Quantum Computing, open to both undergraduate and graduate computer science students. The course follows a lab-integrated learning model that combines conceptual instruction with hands-on programming every week. Students begin with Quantum Without Linear Algebra (QWLA)~\cite{qwla}, a lightweight simulation framework that supports intuitive engagement with core quantum ideas without requiring prior knowledge of matrix algebra. Once students build foundational fluency, the course transitions to IBM Qiskit~\cite{qiskit} for constructing and analyzing quantum circuits and algorithms. Each weekly lab targets a specific learning objective, such as superposition, entanglement, or noise mitigation, and reinforces it through executable experiments and guided reflection.

This instructional approach aims to balance accessibility with intellectual rigor. By grounding theoretical concepts in programming and experimentation, students develop an understanding of quantum program behavior, error sources, and the relationship between abstractions and hardware execution. In this report, we share our experience designing and teaching this lab-integrated course. We describe the motivations behind its instructional design, present the core components of the curriculum, and reflect on student outcomes and feedback. Our goal is to offer a replicable framework that integrates conceptual learning and programming practice in quantum computing education.

\section{Background}

Traditional approaches to teaching quantum computing often emphasize theoretical foundations, typically delivered through physics or mathematics departments. Courses offered in physics programs tend to focus on the behavior of quantum systems and hardware implementation \cite{caltech, berkeley-phys}, while those in mathematics departments emphasize formal aspects of information theory and linear algebra \cite{berkeley-math, waterloo}. In both cases, assessment usually relies on written assignments, exams, and final projects presented as reports or oral presentations on selected algorithms.

These courses are generally targeted at graduate students. Undergraduate quantum computing offerings within computer science departments remain relatively uncommon \cite{ut-austin, ucla}. However, there has been increasing interest in developing programming-oriented and software-focused quantum computing courses that prioritize algorithmic thinking, practical experimentation, and computational abstraction. Notable examples include the University of Washington's course Introduction to Quantum Computing and Quantum Programming in Q\#~\cite{uw, qc-sigcse}, and the work by Seegerer and others~\cite{seegerer}, which explores how core ideas such as superposition, entanglement, and quantum algorithms can be taught through computing education frameworks.

These efforts represent a meaningful shift toward integrating theory with practice. Programming-based instruction has shown that active experimentation can make quantum concepts more accessible to computer science students. Nevertheless, many current courses still lack dedicated laboratory components that allow students to formulate hypotheses, visualize results, and debug their understanding through code. By comparison, classical computing education often begins with hands-on programming, then progresses toward algorithms and theory through iterative practice. Quantum computing education stands to benefit from a similar approach, where theoretical concepts are grounded in coding and experimentation rather than introduced as purely abstract content.
In contrast to prior programming-oriented courses, our work places sustained, weekly laboratory experimentation at the center of conceptual development rather than as a supplement to lectures.

\section{Theoretical Foundation}
The course design was guided by three perspectives:
(1) the core ideas that define conceptual understanding in quantum computing,
(2) a learning process centered on conceptual change through experimentation and reflection, and
(3) a progressive framework that moves from QWLA to Qiskit, linking intuitive understanding with practical programming experience.

\subsection{Core Ideas in QC Education}
Seegerer and others~\cite{seegerer} identify five core ideas that underpin conceptual understanding in quantum computing: superposition, entanglement, quantum computation, quantum algorithms, and quantum cryptography. These ideas define the mental models students must develop to think quantum mechanically. For example, superposition introduces probabilistic reasoning about state, entanglement extends beyond classical correlation, and quantum algorithms demonstrate how interference and measurement can be used to amplify correct outcomes.

Traditional instruction often presents these ideas through dense mathematics or physical analogies, which can obscure their computational relevance. Our course builds on this conceptual framework by embedding each idea within a lab sequence that combines explanation, simulation, and hands-on experimentation. Students first engage with these concepts intuitively, then explore their formal structure through programming exercises that reveal their computational meaning.

\subsection{Conceptual Change and Evolution}
Learning quantum computing involves more than absorbing new content; it requires reconfiguring existing knowledge. Students often bring classical intuitions about computation and probability that conflict with quantum principles. For instance, they may believe that measurement reveals a pre-existing value rather than collapsing a probabilistic state, or assume that correlations imply hidden communication instead of entanglement.

\begin{figure}[t]
  \centering
  \includegraphics[width=0.9\linewidth]{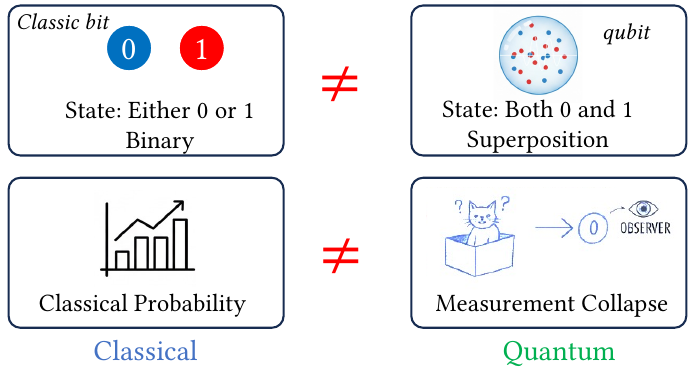}
  \vspace{-1em}
  \caption{Classical versus quantum computational models.}
  \vspace{-2em}
  \label{fig:classic-quantum-conflict}
  \Description{A side-by-side comparison showing how classical bits exist in definite states of 0 or 1, while quantum qubits exist in superposition of both states until measurement collapses the state to a single outcome. This illustrates the fundamental shift from deterministic classical reasoning to probabilistic quantum reasoning that students must develop.}
  \vspace{-1em}
\end{figure}

Figure~\ref{fig:classic-quantum-conflict} illustrates this shift in perspective by comparing classical and quantum models of computation. In classical reasoning, a bit exists in a definite state, either 0 or 1. A qubit can exist in a superposition of both states until it is measured. Measurement collapses the superposition, producing a single outcome. The figure highlights how classical views based on determinism and discrete probability differ from quantum reasoning, where computation depends on amplitudes, interference, and probabilistic results.

The course supports this conceptual evolution by placing students in situations where classical assumptions are challenged through direct experimentation. Each lab provides opportunities for students to observe mismatches between their expectations and actual quantum behavior, which encourages reflection and refinement of their mental models. Experiences such as seeing measurement collapse or encountering non-classical correlations in code serve as turning points that deepen understanding.

This process reflects the broader concept of conceptual change in science education. Students move from initial, often classical, interpretations to more accurate quantum models. 
In our course, this conceptual evolution was operationalized through labs that required students to predict outcomes, run experiments, and reconcile discrepancies between expectation and observation.

\subsection{Progressive Shift to Quantum Programming}
\label{sec:classical-to-quantum-programming}
Teaching quantum programming presents several fundamental challenges that differ from classical programming.
(1) Quantum basis states cannot be directly inspected or debugged, because measurement collapses the state and eliminates traditional feedback mechanisms.
(2) Students must reason across multiple representations, including Dirac notation, matrices, Bloch sphere visualizations, and circuit diagrams, with each offering a distinct perspective on quantum operations.
(3) Entanglement introduces non-local dependencies, which undermine the modular reasoning that is central to classical programming.
(4) Correctness cannot be inferred from single runs. Due to probabilistic outputs and noise, repeated trials and comparison with theoretical expectations are required.

These constraints motivated our decision to introduce quantum programming gradually, using early labs to build intuition before requiring students to reason about full-scale circuits and algorithms.
The course begins with QWLA, which offers an accessible introduction for students with limited backgrounds in quantum mechanics. QWLA enables students to model key phenomena, such as superposition, measurement, and entanglement, using simple Python constructs. This stage builds foundational intuition without requiring advanced mathematics, allowing all students to meaningfully engage with early course content.

Following the introductory phase, the course transitions to IBM Qiskit, which serves as the primary platform for quantum algorithm design, circuit implementation, and experimental analysis. Qiskit offers a professional-grade development environment that enables students to translate conceptual understanding into executable quantum programs. Through systematic comparison of simulation outputs with theoretical predictions, students develop the ability to interpret algorithmic behavior, assess the influence of noise, and reason about computational performance.

This progression from intuitive modeling in QWLA to formal development in Qiskit demonstrates how effective scaffolding can lower entry barriers while guiding students toward conceptual maturity. The integration of core ideas, conceptual evolution, and graduated tooling forms the theoretical foundation of our lab-integrated approach to quantum computing education. 
As discussed in Section~\ref{sec:lessons}, this progression revealed both benefits and limitations.

\section{Instructional Design}
Building on the theoretical foundations described earlier, the instructional design emphasized accessibility, experimentation, and progressive abstraction. The goal was not only to introduce the mechanics of quantum programming but also to help students internalize the reasoning behind quantum phenomena through guided observation and reflection.

\subsection{Learning Approach}
The course followed a lab-integrated learning model in which lectures and labs were closely connected. Each week included one 80-minute lecture and one 80-minute lab session. Lectures introduced key ideas and demonstrated important algorithms and principles. Labs provided guided opportunities for students to apply, test, and extend those ideas. This structure enabled conceptual understanding and computational skills to develop in parallel.

To make the course accessible, instruction began with QWLA, which allowed students to explore foundational concepts such as superposition, measurement, and entanglement using intuitive, code-based representations. 
Initially, quantum states were introduced as Python dictionaries that mapped basis states to amplitudes, providing a simple and concrete way to reason about superposition and measurement without requiring matrix algebra. This representation helped students visualize state transformations and probabilities in a familiar programming context. After students became comfortable with these behaviors, the course transitioned to IBM Qiskit, which served as the main platform for designing algorithms, building circuits, and running experiments. This progression allowed students to move from informal reasoning to formal implementation, connecting conceptual learning with practical skill development.
While this progression was designed to reduce early cognitive load, Sec.~\ref{sec:lessons} reflects on the trade-offs introduced by transitioning between frameworks.

Each lab was explicitly aligned with the concepts introduced in the corresponding lecture. For example, after covering entanglement and measurement correlations, students built and tested Bell-state circuits, analyzed measurement results, and explored the effects of noise and decoherence. These lab activities served as formative checkpoints that helped students connect code and circuits with quantum behavior.

\subsection{Course Structure}
The course was offered over sixteen weeks to both undergraduate and graduate students who had completed an introductory algorithms course. No prior background in quantum mechanics was required. The curriculum (Table~\ref{tab:course-structure}) advanced gradually from foundational concepts to more complex algorithms and applications, balancing theoretical instruction with experimental practice.

\begin{table}[hbt]
\centering
\caption{Weekly lecture topics and laboratory activities}
\vspace{-1em}
\small
\begin{tabular}{p{0.47\linewidth} p{0.47\linewidth}}
\toprule
\textbf{Lecture Topics} & \textbf{Laboratory Focus} \\
\midrule
Qubits, superposition, and measurement & Exploring single-qubit states using QWLA \\
Entanglement and correlations & Building and testing Bell states \\
Phase and interference &  Phase kickback \\
Quantum circuits and the Deutsch-Jozsa algorithm & IBM Qiskit circuits and implementing Deutsch-Jozsa \\
Grover's search & Comparing quantum and classical search performance \\
Noise, and error mitigation & Simulating noise models in Qiskit \\
Quantum Fourier Transform and Phase Estimation & Coding QFT and analyzing frequency outputs \\
Shor's algorithm and applications & Factorization and scaling analysis \\
\bottomrule
\end{tabular}
\label{tab:course-structure}
\Description{A two-column table showing the alignment between weekly lecture topics and corresponding laboratory activities. The table progresses from foundational concepts like qubits and superposition to advanced algorithms like Shor's algorithm, demonstrating the gradual increase in complexity throughout the sixteen-week course.}
\vspace{-1em}
\end{table}

This structure reinforced the idea that each theoretical topic should have a corresponding experimental activity. The gradual increase in complexity helped students build algorithmic thinking alongside scientific reasoning about quantum systems.

\subsection{Assessment and Learning Outcomes}
Assessment combined formative and summative methods to evaluate both conceptual understanding and technical proficiency. Laboratory assignments were assessed automatically using provided testing scripts. The final project required teams of two or three students to design, implement, and evaluate a quantum algorithm or application. Project presentations were evaluated manually based on implementation quality, conceptual insight, and clarity of communication. Example projects included executing Grover's search under noisy conditions and developing small-scale quantum optimization tasks. Additional assessment components included quizzes, a midterm examination, and class participation.

By the end of the course, students were expected to:
\begin{itemize}
    \item Explain core quantum concepts such as superposition, entanglement, and measurement using both descriptive and computational reasoning.
    \item Design and implement quantum circuits and algorithms using Qiskit.
    \item Analyze the effects of noise, error, and decoherence on quantum computation.
    \item Relate theoretical operations to their practical implementations and observable outcomes.
\end{itemize}
The alignment between lectures and labs, along with the gradual progression from intuitive simulation to professional development tools, created a learning environment in which students engaged with quantum computing through experimentation, analysis, and reflection. 


\section{Implementation and Delivery}
The course was delivered over a sixteen-week semester in Fall 2025. It enrolled 9 undergraduate and 6 graduate students in computer science and software engineering. The class met twice weekly, with one 80-minute lecture focused on conceptual foundations and algorithm design, followed by an 80-minute lab session dedicated to hands-on programming and experimentation.

\subsection{Student Background}
\begin{figure}[]
    \centering
    \includegraphics[width=\columnwidth]{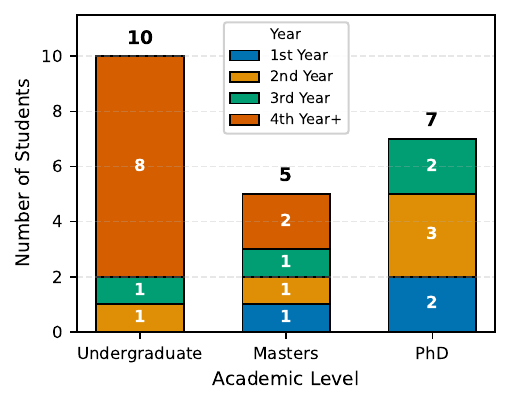}
    \vspace{-3em}
    \caption{Student enrollment by academic level and year.}
    \vspace{-2em}
    \label{fig:student-demographics}
    \Description{A stacked bar chart showing student distribution across three academic levels: Undergraduate, Masters, and PhD. The Undergraduate category shows 8 students in fourth year or beyond and 2 in earlier years. The Masters category shows 5 students, and the PhD category shows 7 students. The chart demonstrates that most enrollees were advanced undergraduates or graduate students.}
\end{figure}

A pre-course survey assessed student preparation across demographics, programming experience, mathematical foundations, and quantum computing familiarity. Most students were senior undergraduates or early-career graduate students seeking exposure to emerging computing paradigms. Figure~\ref{fig:student-demographics} shows the distribution across academic levels, with fourth-year and beyond undergraduates forming the largest cohort.

\begin{figure}[]
    \centering
    \includegraphics[width=\columnwidth]{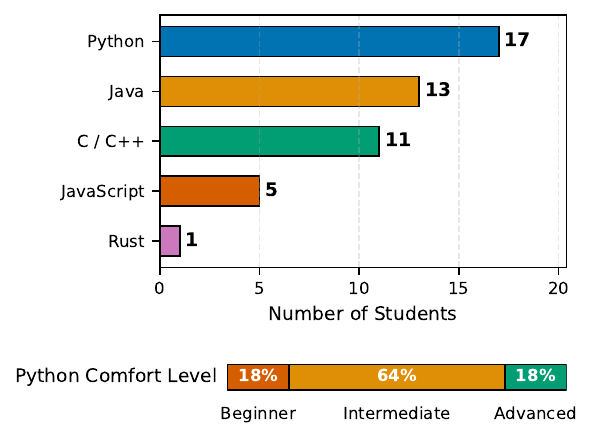}
    \vspace{-3em}
    \caption{Programming language preferences and Python proficiency.}
    \vspace{-2em}
    \label{fig:programming-background}
    \Description{Two horizontal charts showing programming background. The top chart displays absolute counts for preferred languages, with Python most prevalent at 17 students, followed by Java, C and C++, JavaScript, and Rust. The bottom chart shows Python comfort levels as proportions, with 64 percent at intermediate level, 23 percent advanced, and 14 percent beginner.}
\end{figure}

Programming proficiency varied considerably. Python was the most common language, with most students rating their skills as intermediate. Figure~\ref{fig:programming-background} shows language preferences and Python comfort levels. Nearly all students had prior experience with Jupyter Notebooks, which aligned well with the lab-based format.

Mathematical preparation was mixed. Nearly all students had taken linear algebra, though most reported it was from several years earlier. Table~\ref{tbl:math-background} summarizes comfort levels with linear algebra and probability. Most students reported moderate comfort, suggesting that targeted review would be beneficial throughout the course. Almost no students had formal quantum mechanics training beyond introductory physics.

\begin{table}[ht]
\centering
\caption{Mathematical preparation and comfort levels}
\vspace{-1em}
\begin{tabular}{lr}
\hline
Category & Percentage \\
\hline
\textbf{Linear Algebra - Course Taken:} & \\
\quad Yes, recently & 18\% \\
\quad Yes, but long ago & 82\% \\
\quad No & 0\% \\
\textbf{Linear Algebra - Comfort Level:} & \\
\quad Slightly comfortable & 14\% \\
\quad Somewhat comfortable & 41\% \\
\quad Comfortable & 32\% \\
\quad Very Comfortable & 14\% \\
\textbf{Probability \& Statistics - Comfort Level:} & \\
\quad Not comfortable at all & 9\% \\
\quad Slightly comfortable & 9\% \\
\quad Somewhat comfortable & 41\% \\
\quad Comfortable & 32\% \\
\quad Very Comfortable & 9\% \\
\hline
\end{tabular}
\vspace{-1em}
\label{tbl:math-background}
\Description{A table showing mathematical background preparation for 22 students. Most students took linear algebra long ago rather than recently. Comfort levels for both linear algebra and probability cluster around somewhat comfortable and comfortable, with few students at either extreme. This indicates moderate mathematical preparation that would benefit from supplementary review.}
\end{table}

\begin{figure}[]
    \centering
    \includegraphics[width=\columnwidth]{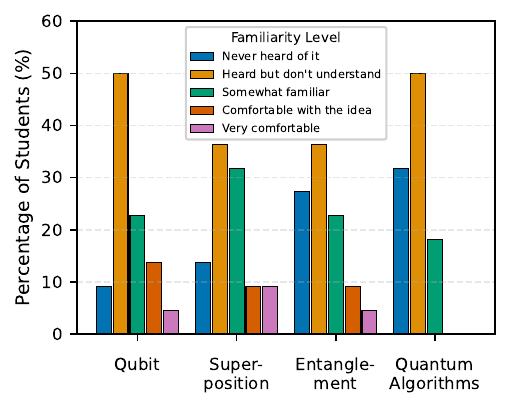}
    \vspace{-3em}
    \caption{Pre-course familiarity with quantum computing concepts.}
    \vspace{-1em}
    \label{fig:quantum-familiarity}
    \Description{A grouped bar chart showing familiarity levels across four quantum concepts: Qubit, Superposition, Entanglement, and Quantum Algorithms. For each concept, five bars represent levels from never heard of it to very comfortable. The most common response across concepts is heard but don't understand, ranging from 36 to 50 percent. Familiarity decreases for more advanced concepts, with Quantum Algorithms showing 32 percent who never heard of it and zero percent very comfortable.}
\end{figure}

Student familiarity with quantum concepts was limited at the start. Figure~\ref{fig:quantum-familiarity} shows awareness decreased for more advanced concepts. Most students had heard of qubits but did not understand them. Familiarity with superposition and entanglement was weaker, and knowledge of quantum algorithms was minimal. This distribution confirmed the need to build algorithmic understanding from foundational principles.

Students expressed primary interest in machine learning applications of quantum computing, followed by optimization and cryptography. Several voiced concerns about mathematical prerequisites and the perceived difficulty of quantum mechanics, which informed our emphasis on computational intuition and hands-on experimentation in early weeks. A teaching assistant provided individualized support during lab sessions to address setup issues and technical questions.

\subsection{Lecture and Laboratory Format}
Lectures introduced theoretical foundations through visualizations and live coding demonstrations. For example, superposition and interference were illustrated with diagrams and probability plots before step-by-step implementation in code. All materials, including slides and annotated Jupyter notebooks, were made available after each session for self-paced review. The format emphasized intuitive understanding before practical implementation and formal mathematics.

Laboratory sessions were designed to extend lecture material through structured experimentation and were informed by prior work on quantum computing education. Early labs employed QWLA to introduce qubits, superposition, and measurement using dictionary-based representations, allowing students to develop intuition without reliance on matrix algebra. Beginning in week four, IBM Qiskit served as the primary platform for circuit construction and algorithm simulation, with an automated testing framework providing immediate feedback on correctness. In line with recommendations by Mykhailova and Svore~\cite{qc-sigcse}, all laboratory sessions included in-person teaching assistant support from an individual with quantum computing experience. Student feedback consistently identified teaching assistant availability as a valuable component of the course, particularly for addressing technical issues and supporting debugging during laboratory exercises.

\subsection{Student Progress and Confidence}
By mid-semester, most students reported feeling somewhat to extremely confident with the material. Understanding of qubits, gates, and measurement was rated well or very well. Entanglement and different representations received slightly lower but still positive ratings. Labs were identified as the most effective component, with students valuing how they reinforced lecture concepts through independent application. Several students noted that creating specific quantum states independently required deep understanding of gate operations.

The most challenging aspects at this stage included creating custom quantum states by combining gates and understanding multi-qubit operations. Technical setup issues consumed significant time in early labs, reducing time for actual problem solving. After the initial learning curve, students found the process more manageable. Some found the custom QWLA API awkward before transitioning to Qiskit, though this early phase proved valuable for building intuition.

Students suggested more examples in slides showing gate operations and their internal workings. They also recommended better integration between lectures and labs, such as dedicating time in the first lecture to review lab logistics and setup. This would allow teaching assistants to focus on conceptual support rather than troubleshooting basic technical issues.

\subsection{Final Projects and Presentations}
The course concluded with team-based final projects where students investigated topics related to quantum algorithms or applications. Example topics included comparing classical and quantum search, analyzing noise mitigation strategies, and implementing small-scale QAOA instances. Each team maintained a GitHub repository with code, documentation, and analysis. Projects were evaluated on implementation correctness, communication clarity, originality, and depth of reflection. The final two sessions featured student presentations demonstrating their work and discussing development challenges.

\subsection{Post-Course Outcomes}
End-of-semester surveys revealed strong understanding of foundational topics. Nearly all students strongly agreed they understood the difference between classical bits and qubits, could describe basic quantum gates, and could explain superposition, entanglement, and measurement outcomes. Students expressed confidence in using both QWLA and IBM Qiskit to build and run quantum circuits.

Understanding of advanced topics showed more variation. Table~\ref{tbl:advanced-topic-understanding} summarizes student confidence across complex concepts. Most students agreed or strongly agreed they understood Grover's algorithm and amplitude amplification. Phase estimation in Shor's algorithm and noise mitigation received more mixed responses, with some indicating only moderate understanding. Understanding of quantum threats to cryptography and post-quantum algorithms was strong overall.

\begin{table}[ht]
\centering
\caption{Understanding of advanced quantum topics}
\vspace{-1em}
\label{tbl:advanced-topic-understanding}
\footnotesize
\begin{tabularx}{\columnwidth}{Xccccc}
\toprule
Topic &
\shortstack{Strongly\\Disagree} &
Disagree &
Neutral &
Agree &
\shortstack{Strongly\\Agree} \\
\midrule
Phase estimation in Shor's & 0\% & 10\% & 10\% & 50\% & 30\% \\
Grover's amplitude amplification & 0\% & 10\% & 30\% & 20\% & 40\% \\
Noise and error mitigation & 0\% & 0\% & 30\% & 40\% & 20\% \\
Quantum threats to RSA/ECC & 0\% & 0\% & 0\% & 40\% & 50\% \\
Post-quantum algorithms & 0\% & 10\% & 10\% & 50\% & 20\% \\
\bottomrule
\end{tabularx}
\Description{A table showing student self-reported understanding of five advanced quantum computing topics. Each row represents a topic with percentage distributions across five agreement levels. Most responses cluster in the agree and strongly agree categories, though phase estimation, noise mitigation, and Grover's algorithm show more neutral responses than cryptography-related topics. No topic received any strongly disagree responses.}
\end{table}

\begin{figure}[ht]
    \centering
    \includegraphics[width=\columnwidth]{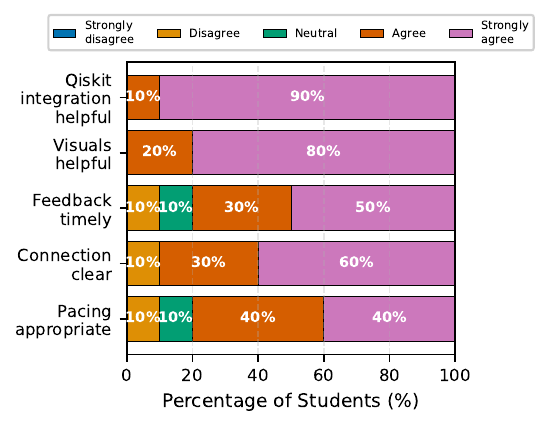}
    \vspace{-3em}
    \caption{Student ratings of course design and delivery.}
    \vspace{-2em}
    \label{fig:course-delivery-ratings}
    \Description{A horizontal stacked bar chart showing satisfaction ratings for five course aspects using a five-point Likert scale. Qiskit integration received 90 percent strongly agree and 10 percent agree. Visual aids received 80 percent strongly agree and 20 percent agree. Connection between materials received 60 percent strongly agree, 30 percent agree, and 10 percent disagree. Timely feedback received 50 percent strongly agree, 30 percent agree, 10 percent neutral, and 10 percent disagree. Appropriate pacing received 40 percent each strongly agree and agree, with 10 percent each neutral and disagree. All aspects show predominantly positive responses with no strongly disagree ratings.}
\end{figure}

Course design and delivery received highly positive ratings. Figure~\ref{fig:course-delivery-ratings} shows student satisfaction across five key aspects. Nearly all students agreed or strongly agreed that pacing was appropriate, connections between materials were clear, and instructor feedback was timely. Visuals and circuit diagrams were particularly valued for aiding understanding. Integration of IBM Qiskit resources received unanimous positive feedback.

The most engaging topics varied by student interest. The Deutsch-Jozsa algorithm was frequently cited for demonstrating quantum speedup and connecting well from lecture to lab. Grover's algorithm appealed to many due to practical applications, though some acknowledged difficulty creating oracles and quantum circuit versions of classical verification logic. Foundational content on qubits and gates was appreciated for establishing groundwork. Transpilation and circuit optimization topics interested students who enjoyed seeing how circuits transformed for execution on different backends. Shor's algorithm engaged students curious about modular exponentiation, period finding, and factor extraction.

The most challenging aspects included cryptography sections, Lab 6 on implementing quantum circuits for function evaluation, and mathematical foundations underlying algorithms. Several students noted a difficulty spike between course halves, with later topics demanding more linear algebra and complex number knowledge. Lab 6 stood out for difficulty in defining functions as quantum circuits and converting binary operations to quantum gates. The first lab was challenging due to limited initial understanding of Qiskit, though it seemed less difficult in retrospect. Final project scoping was challenging, particularly identifying novel contributions with limited subject knowledge.

Students overwhelmingly found hands-on labs extremely useful for learning and reported feeling extremely confident implementing small quantum circuits from scratch. Many expressed pride in mastering the Qiskit library and gaining skills valuable for quantum-related positions. Understanding quantum gates and grasping classical components of Shor's algorithm were identified as significant accomplishments. Several felt confident in fundamentals and capable of independently investigating algorithms beyond course content.

Phase estimation and Fourier transforms remained unclear for some due to reliance on linear algebra not covered deeply. Phase kickback mechanics, particularly why circuits must be reversed, needed more clarity. Aspects of quantum-safe cryptography and Grover's algorithm implementation required further explanation. Most students indicated likelihood to continue exploring quantum computing, and all either recommended the course or might recommend it with reservations.

Key suggestions for improvement included starting with Qiskit earlier rather than using QWLA, better alignment between lectures and labs in early weeks, more foundational linear algebra material on tensor products and state representations, less guided labs requiring more independent problem solving, improved final project guidance with concrete examples and clearer scope expectations, and adjusted pacing to allow more time for advanced topics in the second half.

\subsection{Assessment Results and Performance Analysis}

Assessment combined formative and summative methods, with laboratory assignments contributing the largest portion of the final grade, followed by the final project, an examination, quizzes, and participation. Graduate students performed consistently well across all assessments, with everyone earning A or above. In contrast, undergraduate performance showed greater variation, with final grades ranging from C- to A+. Only one undergraduate student achieved A+, two earned A, and the remaining students received grades between C- and A-. This performance gap reflected differences in mathematical preparation and study habits.

Designing effective laboratory exercises presented notable challenges. Labs needed to be accessible enough for completion within the allocated time while requiring genuine engagement with quantum concepts. To address this, lab exercises provided skeleton code with clearly marked sections for student implementation and inline hints about appropriate quantum operations. This scaffolding achieved high completion rates, but some students felt that the labs were too guided and wished for more independent problem-solving.

Lab 6 proved to be the most challenging one. 
This lab required students to implement quantum oracles for hash function evaluation. Although students understood the conceptual requirements for reversible computation and could explain oracle purposes, translating this understanding into correct gate sequences proved extremely difficult. 
The primary challenge was converting classical boolean logic into reversible quantum circuits. 
While students generally understood the intended behavior of the oracles, they struggled with constructing them using quantum gates.
We noticed an interesting pattern that students often demonstrated solid conceptual understanding in quizzes and explanations, but struggled to translate that understanding into working code. 
This gap between conceptual knowledge and implementation ability appeared across multiple topics, suggesting that quantum programming requires explicit attention to the translation between understanding and practice.

\section{Lessons Learned}
\label{sec:lessons}
The first offering of this course provided several insights into how students engage with quantum computing in a lab-integrated instructional setting. The lessons below are ordered by instructional significance, based on their observed impact on student learning and their implications for future course design. Evidence was drawn from pre-course surveys, mid-semester feedback, end-of-semester surveys, written reflections, assignment outcomes, and instructional observations.

\noindent\textbf{Lesson 1: Conceptual Understanding Does Not Translate Directly to Implementable Quantum Programs.}
The most consequential observation was a persistent gap between students' conceptual understanding of quantum principles and their ability to implement those principles correctly in code. Many students exhibited solid understanding of superposition, entanglement, and measurement in quizzes and written explanations. However, they frequently encountered difficulty when translating this understanding into correct quantum circuits. This gap was particularly evident in tasks involving multi-qubit composition and reversible logic.

These observations indicate that quantum programming involves skills that extend beyond conceptual knowledge. Students must reason at the level of individual gates, track qubit states across circuit execution, manage qubit indexing, decide when measurements are appropriate, and interpret probabilistic outcomes. Such abilities did not emerge automatically from conceptual mastery and therefore require deliberate instructional attention through practice and feedback.

\noindent\textbf{Lesson 2: Oracle Implementation Presents Distinct Challenges.}
Oracle construction emerged as the most challenging programming task in the course. Lab~6, which required students to implement quantum oracles for hash-based evaluation, resulted in lower performance and greater difficulty relative to other labs. Although students generally understood the conceptual role of oracles within quantum algorithms, many struggled to convert classical boolean logic into reversible quantum circuits.

Students often attempted to directly map classical control flow into quantum circuits, resulting in incorrect or non-reversible implementations. Debugging these errors proved difficult, as mistakes frequently became apparent only when the oracle was composed within a larger algorithm such as Grover's search. These observations suggest that oracle implementation introduces conceptual and practical challenges that differ from earlier quantum programming tasks. Future offerings will address this gap by introducing targeted preparatory activities focused on reversible computation and incremental oracle construction prior to full algorithmic integration.

\noindent\textbf{Lesson 3: Transitions Between Programming Frameworks Introduce Cognitive Overhead.}
Beginning the course with QWLA and later transitioning to IBM Qiskit introduced substantial cognitive overhead. While QWLA was intended to reduce early barriers by avoiding matrix algebra, many students reported difficulty adapting to a new programming framework midway through the semester. The transition required relearning syntax, adjusting to new application programming interfaces, and reconciling different representational conventions.

Differences in qubit ordering conventions contributed significantly to confusion. QWLA employed a big-endian representation, whereas Qiskit uses little-endian indexing. Students who developed intuitions about multi-qubit states in QWLA often found that these intuitions did not carry over reliably to Qiskit. Several students suggested beginning directly with Qiskit. Based on this feedback, future offerings will adopt Qiskit from the start and rely on its simulation and visualization features to support early intuition building.

\noindent\textbf{Lesson 4: Scaffolding Within Laboratories Should Be Reduced Over Time.}
Laboratory design required balancing guided instruction with opportunities for independent reasoning. Early labs included substantial scaffolding through skeleton code, inline hints, and step-by-step instructions. This approach supported early progress and helped students acclimate to quantum programming. As the semester progressed, some students reported that this level of guidance limited opportunities for independent problem solving.

Future offerings will reduce within-lab scaffolding over time. Early labs will remain highly structured to support initial skill development. Subsequent labs will provide less embedded guidance and place greater responsibility on students to make design and implementation decisions. Later labs will emphasize synthesis and open-ended problem solving. The final project already followed this approach and showed that students were able to work independently when adequately prepared.

\noindent\textbf{Lesson 5: Experimental Interaction Plays a Central Role in Confidence Development.}
Weekly laboratory experiments contributed substantially to students' confidence and understanding of quantum concepts. Writing small programs, executing them repeatedly, and observing empirical outcomes allowed students to compare expectations with observed behavior. Labs that produced visual or statistical feedback, such as statevector plots and Bell-state measurement distributions, were particularly effective.

This iterative process was especially valuable for students who initially lacked confidence in their ability to engage with quantum mechanics. Continued interaction with executable experiments supported a gradual shift from passive reception to active exploration. At the same time, increased confidence did not consistently translate into success on more complex programming tasks, indicating the need for sustained practice as task complexity increased.

\noindent\textbf{Lesson 6: Course Pacing should Reflect Nonlinear Increases in Cognitive Demand.}
Student feedback indicated that the cognitive demands of the course increased sharply in the later weeks. Early topics, including single-qubit operations and measurement, were generally perceived as manageable. In contrast, later topics involving Grover's algorithm, noise modeling, and Shor's algorithm required greater integration of concepts and more extensive debugging effort. Many students described these topics as compressed and cognitively demanding.

These observations suggest that while scaffolding within individual labs should decrease over time, course-level structural support should increase for implementation-intensive material. Future offerings will compress early material to allocate additional instructional time to advanced algorithms. Planned adjustments include extended timelines for later labs, additional checkpoints, and short preparatory tutorials focused on debugging strategies and algorithmic structure.

\noindent\textbf{Lesson 7: Technical Infrastructure Can Consume Disproportionate Instructional Time.}
Technical setup and environment configuration required more instructional time than anticipated. Students encountered difficulties related to Python environments, Jupyter configuration, package dependencies, and Qiskit installation. Time spent addressing these issues reduced the time available for engaging with quantum concepts and programming tasks.

To limit this overhead, future offerings will prioritize cloud-based Jupyter environments that reduce local configuration requirements. When local installation is unavoidable, a dedicated setup session and detailed configuration guides will be provided before the first programming lab.

\noindent\textbf{Lesson 8: Teaching Assistant Support Is Essential for Technical Troubleshooting.}
The presence of a teaching assistant with quantum computing knowledge during lab sessions proved valuable for student learning. Students identified teaching assistant support as among the most helpful course components. Direct access to technical support during programming activities enabled students to resolve implementation challenges and debug code errors efficiently.

This observation aligns with prior quantum computing courses where Mykhailova and Svore~\cite{qc-sigcse} argued that dedicated programming lab time with in-person access to quantum computing experts greatly improves learning experience. Teaching assistant support was particularly important for addressing technical setup issues, environment configuration problems, and debugging assistance that would otherwise consume significant instructional time.

\paragraph{Summary.}
Taken together, these lessons indicate that lab-integrated instruction can support conceptual development and confidence building in quantum computing education when instructional design choices are aligned with student capabilities. The findings also point to the need for explicit instructional support when moving from conceptual understanding to executable quantum programs.

\section{Conclusions}
This paper reported on the design and evaluation of a lab-integrated quantum computing course that used sustained programming-based experimentation to support conceptual development. Coupling lectures with weekly laboratories helped students build intuition and confidence for foundational quantum concepts, including superposition, measurement, and entanglement. Hands-on interaction with executable quantum programs played a central role in shaping how students reasoned about quantum behavior.

The course also revealed clear limitations of lab-based instruction. A persistent gap emerged between conceptual understanding and correct program implementation, particularly for reversible logic and oracle construction. These findings indicate that quantum programming introduces implementation challenges that require targeted instructional support beyond exploratory experimentation. Instructional design choices related to framework selection, scaffolding, and pacing also substantially influenced student engagement with complex programming tasks.

Future offerings will begin directly with Qiskit to avoid framework transitions, introduce preparatory instruction for reversible computation and oracle implementation, and reduce laboratory scaffolding over time. Course pacing will be adjusted to allocate additional instructional time to implementation-intensive topics in the later part of the semester. 

\balance
\bibliographystyle{ACM-Reference-Format}
\bibliography{biblio}

\end{document}